\documentclass[a4paper]{jpconf}
\usepackage{graphicx}
\usepackage{hyperref}
\usepackage{amssymb}
\usepackage{amsmath}

\newcommand{\ex}[1]{\ensuremath{\times10^{#1}}}
\newcommand{\dmij}[1]{\ensuremath{\Delta m_{#1}^2}}
\newcommand{\Uaj}[1]{\ensuremath{|U_{#1}|^2}}
\newcommand{\nua}[1]{\ensuremath{\rlap{\kern-2.5pt\ensuremath{\overset{\scriptscriptstyle(-)}{\phantom{\nu}}}}{\ensuremath{{\nu}_{#1}}}}}

\begin{document}
\title{Light sterile neutrinos: the current picture from neutrino oscillations}

\author{Stefano Gariazzo}

\address{Instituto de F\'{\i}sica Corpuscular (CSIC-Universitat de Val\`{e}ncia)\\ 
Parc Cient\'{\i}fic UV, C/ Catedr\'atico Jos\'e Beltr\'an, 2, E-46980 Paterna (Valencia), Spain
}

\ead{gariazzo@ific.uv.es}

\begin{abstract}
Light sterile neutrinos with a mass around 1 eV have been studied for many years as a possible explanation of the so called short-baseline neutrino oscillation anomalies.
Recently, several neutrino oscillation experiments reported preferences for non-zero values of the mixing angles and squared mass differences for active-sterile mixing, which however are not always in agreement.
I will review our current knowledge on the light sterile neutrino in the 3+1 model, starting with a separate discussion on the status of the most relevant searches and then analyzing the problems that arise when combining different probes in a global fit.
\end{abstract}

Our current knowledge of the oscillation parameters
in the three neutrino scheme has been improved noticeably in the last twenty years,
see e.g.\ \cite{deSalas:2017kay}.
Yet, several anomalous experimental results remain unexplained \cite{Gariazzo:2015rra}.
The anomalies, discussed in the following,
might have a common explanation if a new neutrino eigenstate exists.
Such scenario is usually labeled ``3+1'' neutrino model
and it proposes
oscillations between the three standard and the fourth, sterile, neutrino,
driven by a new squared mass difference $\dmij{41}=m_4^2-m_1^2\simeq 1$~eV$^2$,
in order to explain the observed anomalies.

Given the above mass splitting,
oscillations between active and sterile neutrinos are dominant
when (see below)
$
\dmij{41} L/E\simeq1
$,
where $L$ and $E$ are the traveled distance and the neutrino energy, respectively.
These parameters define what we call Short BaseLine (SBL) oscillations,
which are not influenced by the three neutrino mixing parameters,
since
the terms corresponding to the solar and atmospheric mass splittings cannot develop at the considered $L/E$.
At SBL, therefore,
only the effect of \dmij{41} must be considered
and
one can write the
transition probability between a neutrino or antineutrino
of flavor $\alpha$ and one of flavor $\beta$
as (see e.g.~\cite{Gariazzo:2015rra}):
\begin{eqnarray}
P^{\mbox{SBL}}_{\nua{\alpha}\rightarrow\nua{\beta}}
&\simeq&
\sin^2 2\vartheta_{\alpha\beta}\sin^2\left(\frac{\dmij{41}L}{4E}\right)
\,,
\qquad(\alpha\neq\beta)
\\
P^{\mbox{SBL}}_{\nua{\alpha}\rightarrow\nua{\alpha}}
&\simeq&
1-
\sin^2 2\vartheta_{\alpha\alpha}\sin^2\left(\frac{\dmij{41}L}{4E}\right)
\,,
\end{eqnarray}
where the effective angles $\vartheta_{\alpha\alpha}$ and $\vartheta_{\alpha\beta}$
depend on the fourth column of the mixing matrix $U$:
\begin{eqnarray}
\sin^22\vartheta_{\alpha\beta}
&=&
4\,\Uaj{\alpha4}\,\Uaj{\beta4}
\,,
\qquad(\alpha\neq\beta)
\\
\sin^22\vartheta_{\alpha\alpha}
&=&
4\,\Uaj{\alpha4}\, (1- \Uaj{\alpha4})
\,.
\end{eqnarray}
Since we generally expect the mixing matrix elements \Uaj{\alpha4} ($\alpha=e,\,\mu,\,\tau$)
to be small,
in order not to alter excessively the phenomenology of three-neutrino oscillations
observed in non-SBL experiments,
we expect the \emph{appearance} effective mixing angles $\vartheta_{\alpha\beta}$ ($\alpha\neq\beta$)
to be quadratically suppressed
with respect to the \emph{disappearance} ones, $\vartheta_{\alpha\alpha}$.

\medskip

The first anomaly in the electron neutrino disappearance channel,
with a statistical significance slightly smaller than $3\sigma$,
was reported by the GALLEX and SAGE experiments \cite{Giunti:2010zu,Kostensalo:2019vmv},
which observed a deficit of electron neutrinos at distances of order 1~m from the source.

In 2011, the new calculation of the electron antineutrino fluxes
from nuclear reactors \cite{Mueller:2011nm,Huber:2011wv}
lead to the discovery of a second anomaly,
coming from a smaller observed event rate in a number of existing neutrino experiments
at reactors \cite{Mention:2011rk}
with respect to the predicted one.
The significance is again $\sim3\sigma$.

A possible explanation for both anomalies is that there may be errors
in the calculation of the unoscillated fluxes.
If the theoretical estimates of the reactor antineutrino flux were wrong,
for instance, the reactor anomaly would also be wrong,
and the same applies to the Gallium case.
Another possibility is that the two anomalies come from a suppression of the measured flux due
to a disappearance generate by a non-zero effective angle $\vartheta_{ee}$.

In order to better investigate a possible neutrino oscillations explanation of these anomalies,
in the recent years several experiments at SBL started to measure the antineutrino flux
at different distances from nuclear reactors.
Such observations can be used to compute flux ratios,
that depend only on neutrino oscillation effects
and not on other systematics that are usually independent of distance,
for example the normalization of the unoscillated flux.
Experiments of this class provide \emph{model-independent} results
because the theoretical model for the unoscillated flux is nearly irrelevant when computing the fit.

The first experiment to provide results obtained with a ratio method was
NEOS \cite{Ko:2016owz}, in South Korea.
A second experiment of this kind is
DANSS \cite{Alekseev:2018efk}, in Russia,
which has a movable detector that can be placed
at three different distances between $\sim10.5$ and $12.5$~m
from the reactor core.
The combined results of NEOS and DANSS,
in 2018,
indicated a preference in favor of
$\dmij{41}\simeq1.3~\mbox{eV}^2$ and
$\Uaj{e4}\simeq0.01$
over the standard three neutrinos case,
with a significance of $\sim3.5\sigma$
\cite{Gariazzo:2018mwd,Dentler:2017tkw}.
Considering the new full dataset by DANSS,
presented for the first time in the EPS-HEP conference in July 2019 \cite{danilov_epshep19},
together with
NEOS \cite{Ko:2016owz}
and
PROSPECT \cite{Ashenfelter:2018iov}
observations,
we obtain a new best-fit point at
$\dmij{41}\simeq 0.4~$eV$^2$ and
$\Uaj{e4}\simeq0.01$,
nearly degenerate with the previous one,
and a reduced model-independent preference in favor of the light sterile neutrino
of $\sim2.5\sigma$
\cite{sterile19}.

\medskip

In the muon (anti)neutrino disappearance channel,
current experiments only provide strong upper bounds
on the matrix element \Uaj{\mu4}, since
no anomaly was ever observed.

Two classes of experiments fall in this category:
atmospheric neutrino oscillation probes,
mainly driven by the
IceCube \cite{Aartsen:2017bap,TheIceCube:2016oqi} observations,
and measurements using accelerator (anti)neutrinos, dominated by the MINOS+ experiment \cite{Adamson:2017uda}.

MINOS+ \cite{Adamson:2017uda}, with its
near ($\sim500$~m from the source)
and far ($\sim800$~km) detectors,
currently provide the strongest bounds on \Uaj{\mu4} over a broad range of \dmij{41} values.
We have verified that the bounds on \Uaj{\mu4}
do not change significantly when the three-neutrino mixing parameters or
the other active-sterile mixing angles are varied
in the analysis \cite{sterile19}.
Given the position of the near detector,
at which for mass splittings $\dmij{41}\gtrsim1$~eV$^2$
active-sterile oscillations can develop,
MINOS+ can use a far-to-near flux ratio to constrain the neutrino mixing in a model-independent way
only for $\dmij{41}\lesssim1$~eV$^2$.
For this reason, in the latest analyses the MINOS+ collaboration decided to use
a full two-detectors fit instead of a ratio fit,
although
in the high \dmij{41} range the bounds have a significant dependence on cross-section systematics.
We have checked that, in the most interesting region below $\dmij{41}\lesssim10$~eV$^2$,
a far-to-near ratio analysis gives results very similar
to those obtained with the full two-detectors fit \cite{sterile19}.

\medskip

The  LSND \cite{Aguilar:2001ty}
and
MiniBooNE \cite{Aguilar-Arevalo:2018gpe}
(anti)neutrino appearance experiments
are responsible for the
most controversial anomalies in SBL oscillations until now.

The LSND experiment,
considering a beam of muon antineutrinos,
was the first one to report the anomalous appearance of electron antineutrinos,
with a significance of $\sim3.8\sigma$.
The KARMEN experiment,
working at slightly smaller distances,
never confirmed the anomaly \cite{Armbruster:2002mp}.

The MiniBooNE experiment, built to test the LSND anomaly,
uses neutrinos at higher energies,
preserving approximately the same $L/E$.
The most recent MiniBooNE results \cite{Aguilar-Arevalo:2018gpe}
are in partial agreement with the LSND ones.
The preferred best-fit by MiniBooNE,
corresponding to maximal mixing between active and sterile states,
is however in tension with the
ICARUS \cite{Antonello:2013gut}
and
OPERA \cite{Agafonova:2013xsk}
results,
and moreover it is not really sufficient to fully explain the excess
in the two bins at the lowest studied energies.
For these reasons, a new experiment, MicroBooNE \cite{Chen:2007ae},
was proposed to check the LSND and MiniBooNE excess.
MicroBooNE uses liquid Argon time projection chamber (LArTPC) technology
with the aim of being able to achieve
a better level of signal/background separation.
This should allow us to determine if the anomalous events are really due to neutrino oscillations
or to some other kind of new physics.

\medskip

The effective angles entering
electron (anti)neutrino disappearance ($\vartheta_{ee}$),
muon (anti)neutrino disappearance ($\vartheta_{\mu\mu}$)
and 
electron (anti)neutrino appearance ($\vartheta_{e\mu}$)
oscillation formulas
can be written in terms of two elements of the fourth column of the neutrino mixing matrix:
\Uaj{e4} and \Uaj{\mu4}.
When combining appearance and disappearance data
in a global fit, we constrain such matrix elements.
From the model-independent fit of NEOS and DANSS data we obtain
a $3\sigma$ upper limit $\Uaj{e4}\lesssim3\ex{-2}$.
From the muon disappearance channel,
mainly driven by MINOS+ and IceCube,
we have a $3\sigma$ upper bound
$\Uaj{\mu4}\lesssim10^{-2}$.
The combined bound from disappearance probes
is therefore
$\sin^22\vartheta_{e\mu}=4\Uaj{e4}\Uaj{\mu4}\lesssim10^{-3}$ at $3\sigma$,
but the LSND and MiniBooNE anomalies
require a mixing angle
$\sin^22\vartheta_{e\mu}\gtrsim10^{-3}$, again at $3\sigma$.
We do not need further details to see that
there is a tension between appearance and disappearance observations.

For quantifying the tension between the two sets of constraints,
adopting a parameter goodness of fit (PG) test
on the best-fit point is the easiest way.
The $p$-value of the PG for the full combination of appearance and disappearance data,
including the most recent results,
is around $10^{-9}$ \cite{sterile19},
certainly too small to be due to random realizations
of the same underlying model.
This indicates that there is no common sterile neutrino solution for the SBL anomalies.
In order to reconcile appearance and disappearance probes some additional explanation is required.

Using the PG, one can test which experiment is mostly responsible
for the tension \cite{sterile19,Dentler:2018sju}.
Assuming that the model-independent observations of NEOS and DANSS
are not influenced by unaccounted systematics or new physics,
and since the muon disappearance experiments observe no anomaly,
we are left with questioning the effects of LSND, MiniBooNE or both.
When we perform the global fit excluding MiniBooNE,
which alone has a $4.8\sigma$ preference in favor of a sterile neutrino,
the $p$-value becomes close to $10^{-6}$:
significantly larger, but not enough to solve the tension.
On the other hand, if we remove LSND,
with its $3.8\sigma$ preference in favor of the 3+1 neutrinos case,
we obtain a $p$-value of approximately $10^{-5}$:
an order of magnitude larger than in the case without MiniBooNE.
From these numbers we learn that
each experiment alone can quote
a preference for the 3+1 model which
does not reflect its role in the global fit.
LSND has a bigger effect on the global analysis because its best-fit
is not as much in tension with other experiments
as the MiniBooNE one.
Only if all the other data are ignored,
therefore, one can claim that
MiniBooNE currently gives the strongest preference in favor of the 3+1 scenario.

If we finally remove both LSND and MiniBooNE from the global analysis,
we are left with no anomalous signal in the appearance channel.
The tension vanishes and the remaining experiments give a consistent fit
where \Uaj{\mu4} is compatible with zero and \Uaj{e4} is given by reactor experiments.
Such analysis makes sense if there is new physics beyond the light sterile neutrino:
if additional new physics is responsible for the LSND and MiniBooNE anomalies,
it is incorrect to include their data in a global fit of the 3+1 mixing parameters.

\medskip

In the incoming months, many experiments will publish more results.
Among the ones that are expected to have a significant impact,
we have
STEREO \cite{Almazan:2018wln,Bernard:2019jli}
and
PROSPECT \cite{Ashenfelter:2018iov},
whose current limits are not competitive enough to confirm or reject
the best-fit by DANSS and NEOS,
but they will reach soon the required sensitivity.
Within the next few years, with more data, we will either have
a strong preference in favor of a common preferred point
or
a final rejection of the
light sterile neutrino explanation of the anomalies.
In the former case,
with many experiments independently observing oscillations that involve a new neutrino state
and the same mixing parameters,
we will have the cleanest signal ever observed in favor of new physics beyond the standard model.
In any case, the already mentioned
MicroBooNE \cite{Chen:2007ae}
experiment,
apart for giving a final confirmation or disproval
of the sterile neutrino interpretation of the LSND and MiniBooNE results,
is also expected to indicate us
if the anomalies can be due to some other kind of new physics, whatever it is.

All together, these new experiments will drive us towards a deeper understanding
of the SBL anomalies and of the new physics that potentially produces them.
If a consistent explanation in terms of a light sterile neutrino is viable,
moreover,
we will finally know
which are the mixing parameters associated to it.

\section*{Acknowledgements}
The author receives support from the European Union's Horizon 2020 research and innovation programme under the Marie Sk{\l}odowska-Curie individual grant agreement No.\ 796941.

\section*{References}
\bibliography{main}

\end{document}